\documentclass[11pt]{amsart}

\usepackage{amscd}
\usepackage{amsmath}
\usepackage{graphicx}
\usepackage{amsfonts}
\usepackage{amssymb}
\begin{document}

\title[Optimal entanglement witnesses]
{Optimality of a class of entanglement witnesses for $3\otimes 3$
systems}

\author{Xiaofei Qi}
\address[Xiaofei Qi]{
Department of Mathematics, Shanxi University , Taiyuan 030006, P. R.
 China;} \email{xiaofeiqisxu@yahoo.com.cn}

\author{Jinchuan Hou}
\address[Jinchuan Hou]{Department of
Mathematics\\
Taiyuan University of Technology\\
 Taiyuan 030024,
  P. R.  China}
\email{jinchuanhou@yahoo.com.cn}


\thanks{This work is partially supported by a grant from
International Cooperation Program in Sciences and Technology  of
Shanxi (2011081039),  Natural Science Foundation of China (11171249,
11101250) and Youth Foundation of Shanxi Province (2012021004).}

\begin{abstract}

Let $\Phi_{t,\pi}: M_3({\mathbb C}) \rightarrow M_3({\mathbb C})$ be
a linear map defined by
$\Phi_{t,\pi}(A)=(3-t)\sum_{i=1}^3E_{ii}AE_{ii}+t\sum_{i=1}^3E_{i,\pi(i)}AE_{i,\pi(i)}^\dag-A$,
where   $0\leq t\leq 3$ and $\pi$ is a permutation of $(1,2,3)$. We
show that the   Hermitian matrix $W_{\Phi_{t,\pi}}$ induced by
$\Phi_{t,\pi}$ is an optimal  entanglement witness if and only if
$t=1$ and $\pi$ is cyclic.

\end{abstract}
\maketitle

\section{Introduction}

Let $H$  be a separable complex Hilbert space. Recall that a quantum
state on $H$ is a density operator $\rho\in{\mathcal B}(H)$ which is
positive and has trace 1. Denote by ${\mathcal S}(H)$ the set of all
states on $H$. If $H$ and $K$ are finite dimensional, a state in the
bipartite composition system  $\rho\in{\mathcal S}(H\otimes K)$ is
said to be separable if $\rho$ can be written as $\rho=\sum_{i=1}^k
p_i \rho_i\otimes \sigma _i,$ where $\rho_i$ and $\sigma_i$ are
states on $H$ and $K$ respectively, and $p_i$ are positive numbers
with $\sum _{i=1}^kp_i=1$. Otherwise, $\rho$ is entangled.

Entanglement is an important physical resource to realize various
quantum information  and quantum communication tasks such as
teleportation, dense coding, quantum cryptography and key
distribution \cite{LK,NC}. It is very important but also difficult
to determine whether or not a state in a composite system is
separable. One of the most general approaches to characterize
quantum entanglement  for bipartite composition systems is based on
the notion of entanglement witnesses (see \cite{Hor}). A Hermitian
matrix $W$ acting on $H\otimes K$ is  an entanglement witness
(briefly, EW) if $W$ is not positive and ${\rm Tr}(W\sigma)\geq 0$
holds for all separable states $\sigma$. Thus, if $W$ is an EW, then
there exists an entangled state $\rho$ such that ${\rm Tr}(W\rho)<0$
(that is, the entanglement of $\rho$ can be detected by $W$). It was
shown that, a state is entangled if and only if it is detected by
some entanglement witness \cite{Hor}. Constructing entanglement
witnesses is a hard task, too. There was a considerable effort in
constructing and analyzing the structure of entanglement witnesses
\cite{B,CK, HQ,JB, TG}. However, complete characterization and
classification of EWs is far from satisfactory.

Due to the Choi-Jamio{\l}kowski isomorphism \cite{C,J}, a Hermitian
matrix $W\in{\mathcal B}(H\otimes K)$ with $\dim H\otimes K<\infty$
is an EW if and only if there exists a positive linear map which is
not completely positive (NCP) $\Phi:{\mathcal
B}(H)\rightarrow{\mathcal B}(K)$ and a maximally entangled state
$P^+\in{\mathcal B}(H\otimes H)$ such that $W=W_\Phi=(I_n\otimes
\Phi)P^+$. Recall that a maximally entangled state is a pure state
$P^+=|\psi^+\rangle\langle\psi^+|$ with
$|\psi^+\rangle=\frac{1}{\sqrt{n}}(|11\rangle+|22\rangle+\cdots
|nn\rangle) $, where $n=\dim H$ and $\{|i\rangle\}_{i=1}^n$ is an
orthonormal basis of $H$. Thus, up to a multiple by positive scalar,
$W_\Phi$ can be written as the matrix
$W_\Phi=(\Phi(E_{ij}))_{n\times n}$, where  $E_{ij}=|i\rangle\langle
j|$. For a positive linear map $\Phi:{\mathcal
B}(H)\rightarrow{\mathcal B}(K)$, we always denote $W_\Phi$  the
Choi-Jamio{\l}kowski matrix of $\Phi$ with respect to a given basis
of $H$, that is $W_\Phi=(\Phi(E_{ij}))_{n\times n}$, and we say that
$W_\Phi$ is the witness induced by the positive map $\Phi$.
Conversely, for an EW $W$, we denote $\Phi_W$ for the associated
positive map so that $W=W_{\Phi_W}$.

For any entanglement witness $W$,  let
$\mathcal{D}_W=\{\rho:\rho\in{\mathcal S}(H\otimes K), {\rm
Tr}(W\rho)<0\},$ that is, ${\mathcal D}_W$ is the set of all
entangled states that detected by $W$. For entanglement witnesses
$W_1,W_2$, we say that $W_1$ is finer than $W_2$ if
$\mathcal{D}_{W_2}\subset \mathcal{D}_{W_1}$, denoted by $W_2\prec
W_1$. While, an entanglement witness $W$ is optimal  if there exists
no other witness finer than it. Obviously, a state $\rho$ is
entangled if and only if there is some optimal EW such that ${\rm
Tr}(W\rho)<0$. In \cite{LK}, Lewenstein, Kraus, Cirac and Horodecki
proved that: (1) $W$ is an optimal entanglement witness if and only
if $W-Q$ is no longer an entanglement witness for arbitrary positive
operator $Q$; (2) $W$ is optimal if ${\mathcal P}_W=
\{|e,f\rangle\in H\otimes K: \langle e, f|W|e,f\rangle=0\}$ spans
the whole $H\otimes K$ (in this case, we say that $W$ has spanning
property).  However, the criterion (2) is only a sufficient
condition. There are known optimal witnesses that have no spanning
property, for example, the entanglement witnesses induced by the
Choi maps. Recently, Qi and Hou in \cite{QH} gave a necessary and
sufficient condition for the optimality of entanglement witnesses in
terms of positive linear maps.

{\bf Theorem 1.1.} (\cite[Theorem 2.2]{QH}) {\it Let $H$ and $K$ be
finite dimensional complex Hilbert spaces. Let $\Phi:{\mathcal
B}(H)\rightarrow {\mathcal B}(K)$ be a positive linear map. Then
$W_\Phi$ is an optimal entanglement witness if and only if, for any
$C\in {\mathcal B}(H,K)$, the map $X\mapsto \Phi(X)-CXC^\dag$ is not
a positive map.}

This approach is practical for some situations, especially when the
witnesses have no spanning property. Applying it, Qi and Hou
\cite{QH} showed that the entanglement witnesses arising from some
positive maps in \cite{QH1} are indecomposable optimal witnesses.

If $\dim H=n$, by fixing an orthonormal basis, one may identify
${\mathcal B}(H)$ with $M_n({\mathbb C})$, the $n\times n$ complex
matrix algebra. In this note, we will consider the linear maps
$\Phi_{t,\pi}$ defined by
$$\Phi_{t,\pi}(X)={\small\left(\begin{array}{ccc}(2-t)x_{11}+tx_{\pi(1),\pi(1)}&-x_{12}&-x_{13}\\
-x_{21}&(2-t)x_{22}+tx_{\pi(2),\pi(2)}&-x_{23}\\
-x_{31}&-x_{32}&(2-t)x_{33}+tx_{\pi(3),\pi(3)}\end{array}\right)},\eqno(1.1)$$
where $X=(x_{ij})\in M_3({\mathbb C})$, $0\leq t\leq 3$ and $\pi$ is
any permutation of $(1,2,3)$. We will show that the necessary and
sufficient condition for the Hermitian matrix $W_{\Phi_{t,\pi}}$ to
be an optimal entanglement witness is that $t=1$ and $\pi$ is cyclic
(Theorem 2.2).

\section{Main result and   proof}

In this section, we give the main result and its proof.

Let $\pi$ be a permutation of $(1,2,\ldots,n)$ and $0\leq t\leq n$.
For a subset $F$  of $\{1,2,\ldots,n\}$, if $\pi(F)=F$, we say $F$
is an invariant subset of $\pi$.  Let  $F$ be an invariant subset of
$\pi$. If both $G\subseteq F$ and $G$ is invariant under $\pi$ imply
$G=F$, we say $F$ is a minimal invariant subset of $\pi$. It is
obvious that a minimal invariant subset is a loop of $\pi$ and
$\{1,2,\ldots, n\}=\cup_{s=1}^r F_s$, where $\{F_s\}_{s=1}^r$ is the
set of all minimal invariant subsets of $\pi$. Denote by $\# F_s$
the cardinal number of $F_s$. Then $\sum_{s=1}^r \# F_s=n$. We call
$\max\{\# F_s: s=1,2,\ldots,r\}$ the length of $\pi$, denoted by
$l(\pi)$. In the case that $l(\pi)=n$, we say that $\pi$ is cyclic.

The following lemma was shown in  \cite{HLPQS}.

{\bf Lemma 2.1.}   {\it For any permutation $\pi$ of $\{1,2,3\}$,
let $\Phi_{t,\pi}: M_3({\mathbb C}) \rightarrow M_3({\mathbb C})$ be
a map defined by Eq.(1.1). Then $\Phi_{t,\pi}$ is positive if and
only if $0\leq t\leq \frac{3}{l(\pi)}$.}

The following is our main result in this note, which states that
$W_{\Phi_{t,\pi}}$ is an optimal EW if and only if $t=1$ and $\pi$
is cyclic.

{\bf Theorem 2.2.} {\it For any permutation $\pi$ of $\{1,2,3\}$,
let $\Phi_{t,\pi}: M_3({\mathbb C}) \rightarrow M_3({\mathbb C})$ be
the  map defined by Eq.(1.1).   Then $W_{\Phi_{t,\pi}}$ is an
optimal entanglement witness if and only if $t=1$ and $l(\pi)=3$.}

Before stating the main results in this section, let us recall some
notions and give two lemmas that we  needed.

Let $l$, $k\in\mathbb{N}$ (the set of all natural numbers),  and let
$A_{1},\cdots, A_{k}$, and $C_{1},\cdots, C_{l}\in {\mathcal B}(H$,
$K$). If, for each $|\psi\rangle\in H$, there exists an $l\times k$
complex matrix $(\alpha _{ij}(|\psi\rangle))$ (depending on
$|\psi\rangle$) such that
$$
C_{i}|\psi\rangle=\sum _{j=1}^{k}\alpha
_{ij}(|\psi\rangle)A_{j}|\psi\rangle,\qquad i=1,2,\cdots ,l,
$$
we say that $(C_{1},\cdots ,C_{l})$ is a locally linear combination
of $(A_{1},\cdots ,A_{k})$, $(\alpha_{ij}(|\psi\rangle))$ is called
a {\it local coefficient matrix} at $|\psi\rangle$.  Furthermore, if
a local coefficient matrix $(\alpha_{ij}(|\psi\rangle))$
  can be chosen for every $|\psi\rangle\in H$ so that
its operator norm $\|(\alpha _{ij}(|\psi\rangle))\|=\sup\{\|(\alpha
_{ij}(|\psi\rangle)|x\rangle\|: |x\rangle\in {\mathbb C}^{\rm k},
\||x\rangle\|\leq 1\}\leq 1$, we say that $(C_{1},\cdots ,C_{l})$ is
a {\it contractive locally linear combination} of $(A_{1},\cdots
,A_{k})$; if there is a matrix $(\alpha_{ij})$ such that $C_{i}=\sum
_{j=1}^{k}\alpha _{ij}A_{j}$ for all $i$, we say that $(C_{1},\cdots
,C_{l})$ is a {\it linear combination} of $(A_{1},\cdots ,A_{k})$
with coefficient matrix $(\alpha _{ij})$.

The following characterization of  positive linear maps was obtained
in \cite{H4}, also, see \cite{H}.

\textbf{Lemma 2.3.}  {\it Let $H$ and $K$ be complex Hilbert spaces
of any dimension, $\Phi: {\mathcal B}(H)\rightarrow{\mathcal B}(K)$
be a linear map defined by $\Phi(X) =\sum
_{i=1}^{k}C_{i}XC_{i}^{\dagger}-\sum
_{j=1}^{l}D_{j}XD_{j}^{\dagger}$ for all $X$. Then $\Phi $ is
positive if and only if $(D_{1},\cdots ,D_{l})$ is a contractive
locally linear combination of $(C_{1},\cdots ,C_{k})$. Furthermore,
$\Phi$  is completely positive if and only if $(D_{1},\cdots
,D_{l})$ is a linear combination of $(C_{1},\cdots ,C_{k})$ with a
contractive coefficient matrix, and in turn, if and only if there
exist $E_1, E_2, \ldots , E_r$  in ${\rm span}\{C_{1},\cdots
,C_{k}\}$  such that $ \Phi=\sum_{i=1}^r E_i(\cdot )E_i^\dagger.$}

{\bf Lemma 2.4.} {\it Let $t$ be a fixed number with $0<t<1$ and let
$x_1,x_2,x_3$ be any positive numbers with $x_1x_2x_3=1$ and
$(x_1,x_2,x_3)\not=(1,1,1)$. Then we have
$$\frac{1-\sum_{i=1}^3\frac{1}{(3-t)+tx_i}}{\sum_{i=1}^2\frac{1}{(3-t)+tx_i}
-\frac{4}{(3-t+tx_1)(3-t+tx_2)}-\frac{1}{(3-t+tx_1)(3-t+tx_3)}-\frac{1}{(3-t+tx_2)(3-t+tx_3)}}\geq
(1-t). $$ }

{\bf Proof.} Let $f$ be  the function   in $3$-variables defined
 by
$$\begin{array}{rl}&f(x_1,x_2,x_3)\\
=&\frac{1-\sum_{i=1}^3\frac{1}{(3-t)+tx_i}}{\sum_{i=1}^2\frac{1}{(3-t)+tx_i}
-\frac{4}{(3-t+tx_1)(3-t+tx_2)}-\frac{1}{(3-t+tx_1)(3-t+tx_3)}-\frac{1}{(3-t+tx_2)(3-t+tx_3)}},\end{array}$$
where $t$ is fixed with $0<t<1$ and  $x_1,x_2,x_3$ are any positive
numbers with $x_1x_2x_3=1$ and $(x_1,x_2,x_3)\not=(1,1,1)$. Since
the denominator of $f(x_1,x_2,x_3)$ is not zero whenever
$(x_1,x_2,x_3)\not=(1,1,1)$, a computation shows that
$$\begin{array}{rl}&f(x_1,x_2,x_3)\geq (1-t)\\
\Leftrightarrow&1-\sum_{i=1}^3\frac{1}{(3-t)+tx_i}\geq
(\sum_{i=1}^2\frac{1}{(3-t)+tx_i}
-\frac{4}{(3-t+tx_1)(3-t+tx_2)}\\
&-\frac{1}{(3-t+tx_1)(3-t+tx_3)}-\frac{1}{(3-t+tx_2)(3-t+tx_3)})(1-t)\\
\Leftrightarrow &g(x_1,x_2,x_3)\geq 0,\end{array}$$ where
$$\begin{array}{rl}g(x_1,x_2,x_3)=&(2t^2-2t-3)+(1-t)x_1+(1-t)x_2+(1-t^2)x_3\\
&+(2t-t^2)x_1x_2+tx_2x_3+tx_1x_3.\end{array}$$ Thus,  to complete
the proof of the lemma, we only need to  check that  the minimum of
the $3$-variable function $g$ is zero on the region $x_{i}>0$ with
$x_{1}x_{2}x_3=1$, $i=1,2,3$.

To do this, let
$$L(x_1,x_2,x_3,\lambda)=g(x_1,x_2,x_3)+\lambda(x_{1}x_{2}x_3-1).$$ By the method of
Lagrange multipliers, we have the system
$$\begin{cases}L_{x_1}^\prime=(1-t)+(2t-t^2)x_2+tx_3+\lambda x_2x_3=0,\\
L_{x_2}^\prime=(1-t)+(2t-t^2)x_1+tx_3+\lambda x_1x_3=0,\\
L_{x_3}^\prime=(1-t^2)+tx_2+tx_1+\lambda x_1x_2=0,\\
L_\lambda^\prime=x_1x_2x_3-1=0.\end{cases}\eqno(2.1)$$ Solving this
system, one obtains $$(x_2-x_1)(2t-t^2+\lambda x_3)=0,$$ which
implies that $${\rm either}\ \ \ \ x_1=x_2\ \ \ \ {\rm or}\ \ \ \
2t-t^2+\lambda x_3=0.$$

If $2t-t^2+\lambda x_3=0$, by Eq.(2.1), one   gets
$x_3=\frac{t-1}{t}<0$, a contradiction. Hence we must have
$x_1=x_2$. Thus, by Eq.(2.1) again, we have
$$(2t-t^2)x_1^4+(1-t)x_1^3-tx_1+(t^2-1)=0,$$that is,
$$(x_1-1)[(2t-t^2)x_1^3+(1+t-t^2)x_1^2+(1+t-t^2)x_1+(1-t^2)]=0.\eqno(2.2)$$
Note that $(2t-t^2)x_1^3+(1+t-t^2)x_1^2+(1+t-t^2)x_1+(1-t^2)>0$ for
all $x_1>0$ and $0<t<1$. So Eq.(2.2) holds if and only if  $x_1=1$,
which forces $x_2=x_3=1$. It follows that the function
$g(x_1,x_2,x_3)$ takes its extremum at the point $(1,1,1)$.
Moreover, it is easy to check that $(1,1,1)$ is the minimal point of
$g(x_1,x_2,x_3)$. Hence $g(x_1,x_2,x_3)\geq g(1,1,1)=0$ for all
$x_{i}>0$ with $x_{1}x_{2}x_3=1$, $i=1,2,3$.

Therefore, the inequality in Lemma 2.4 holds for all $x_i>0$,
$i=1,2,3$, with $x_1x_2x_3=1$ and $(x_1,x_2,x_3)\not=(1,1,1)$. The
proof is finished. \hfill$\Box$

Now we are in a position to give the proof of Theorem 2.1.

{\bf Proof of Theorem 2.1.} By Lemma 2.1, $\Phi_{t,\pi}$ is positive
whenever $0\leq t\leq \frac{3}{l(\pi)}$. We will prove the theorem
by considering several cases. Note that, $\Phi_{0,\pi}$ is
completely positive; so we may assume that $t>0$.

{\bf Case 1.} $l(\pi)=1$.

if $l=1$, then $\pi={\rm id}$ (the identical permutation). In this
case, $\Phi_{t,\pi}$ is a completely positive linear map for all
$0<t\leq 3$ (see \cite[Proposition 2.7]{QH1}), and so
$W_{\Phi_{t,\pi}}\geq 0$, which is not an EW.

{\bf Case 2.} $l(\pi)=2$.

If $l=2$, then $\pi^2={\rm id}$. Without loss of generality, assume
that $\pi(1)=2$, $\pi(2)=1$ and $\pi(3)=3$. Since
$\Phi_{t,\pi}(E_{11})=(2-t)E_{11}+tE_{22}$,
$\Phi_{t,\pi}(E_{22})=(2-t)E_{22}+tE_{11}$,
$\Phi_{t,\pi}(E_{33})=2E_{33}$ and $\Phi_{t,\pi}(E_{ij})=-E_{ij}$
with $1\leq i\not=j\leq 3$, the Choi matrix of $\Phi_{t,\pi}$ is
$$\begin{array}{rl}W_{\Phi_{t,\pi}}=&\sum_{i=1}^3(2-t)E_{ii}\otimes
E_{ii}+tE_{22}\otimes E_{11}+tE_{11}\otimes E_{22}+tE_{33}\otimes
E_{33}-\sum_{i\not=j}E_{ij}\otimes E_{ij}\\
=&(2-t)E_{11}\otimes E_{11}+(2-t)E_{22}\otimes E_{22}+2E_{33}\otimes
E_{33}\\&+tE_{22}\otimes E_{11}+tE_{11}\otimes
E_{22}-\sum_{i\not=j}E_{ij}\otimes E_{ij}.\end{array}$$

If $1\leq t\leq\frac{3}{2}$, then let
$$C_1=(2-t)E_{11}\otimes E_{11}+(2-t)E_{22}\otimes E_{22}+2E_{33}\otimes
E_{33}-\sum_{i\neq j; \pi(i)\neq j}E_{ij}\otimes E_{ij}$$and
$$C_2=tE_{22}\otimes E_{11}+tE_{11}\otimes
E_{22}-E_{12}\otimes E_{12}-E_{21}\otimes E_{21}.$$ It is easily
checked that $C_1\geq 0$. As $C_2^{{\rm T}_2}=tE_{22}\otimes
E_{11}+tE_{11}\otimes E_{22}-E_{12}\otimes E_{21}-E_{21}\otimes
E_{12}\geq 0,$ we see that $C_2$ is PPT. It is clear that
$C_1\not=0$ and  $W_{\Phi_{t,\pi}}=C_1+C_2$. Hence
$W_{\Phi_{t,\pi}}$ is decomposable and  not optimal.

If $0<t<1$, then let
$$\begin{array}{rl}D_1=&(2-t)E_{11}\otimes E_{11}+(2-t)E_{22}\otimes E_{22}+2E_{33}\otimes
E_{33}\\&-\sum_{i\neq j; \pi(i)\neq j}E_{ij}\otimes
E_{ij}-(1-t)E_{12}\otimes E_{12}-(1-t)E_{21}\otimes
E_{21}\end{array}$$and
$$D_2=tE_{22}\otimes E_{11}+tE_{11}\otimes
E_{22}-tE_{12}\otimes E_{12}-tE_{21}\otimes E_{21}.$$ It is also
clear that $D_2$ is PPT and $D_1\geq 0$. We still have $D_1\not=0$
and $W_{\Phi_{t,\pi}}=D_1+D_2$. Hence $W_{\Phi_{t,\pi}}$ is
decomposable and    not optimal.

{\bf Case 3.} $l(\pi)=3$, i.e., $\pi$ is cyclic.

If $l(\pi)=3$ and $t=1$, then $\pi$ is a cyclic permutation, and by
\cite[Theorem 3.2]{QH1}, $W_{\Phi_{1,\pi}}$ is optimal.

In the sequel we always assume that $l(\pi)=3$. Our aim is to prove
that $W_{\Phi_{t,\pi}}$ is not optimal for any $0<t<1$. Without loss
of generality, let $\pi(i)=(i+1)\ {\rm mod}\ 3$, $i=1,2,3$. By
Theorem 1.1, to prove that $W_{\Phi_{t,\pi}}$ is not optimal, we
have to prove that there exists a matrix   $C\in M_3({\mathbb C})$
such that the linear map $A\mapsto\Phi_{t,\pi}(A)-CAC^\dagger$ is
positive. Indeed, we will show that, for any positive number
$0<c\leq\sqrt{1-t}$, let $C_0={\rm diag}(c,-c,0)$; then the map
$A\mapsto\Phi_{t,\pi}(A)-CAC^\dagger$ is positive.

To do this, let $C_0={\rm diag}(c,-c,0)$ with $c>0$ and let
$\Psi_{C_0}$ be the map defined by
$$\begin{array}{rl}\Psi_{C_0}(A)=&\Phi_{t,\pi}(A)-C_0AC_0^\dag\\
=&(3-t)\sum_{i=1}^3E_{ii}AE_{ii}^\dagger
+\sum_{i=1}^{3}E_{i,i+1}AE_{i,i+1}^\dagger-A-C_0AC_0^\dag\end{array}$$for
all $A\in M_3({\mathbb C})$.

If $\Psi_{C_0}$ is positive, then by Lemma 2.3, for any unit
$|x\rangle\in {\mathbb C}^3$, there exist scalars $\{
\alpha_{i}(|x\rangle)\}_{i=1}^{3}$, $\{
\beta_{i}(|x\rangle)\}_{i=1}^{3}$, $\{\delta_i(|x\rangle)\}_{i=1}^3$
and $\{\gamma_i(|x\rangle)\}_{i=1}^3$ such that
$$|x\rangle=I|x\rangle=\sum_{i=1}^3\alpha_i(|x\rangle)(\sqrt{3-t}E_{ii})|x\rangle+\sum_{i=1}^3\beta_i(|x\rangle)\sqrt{t}E_{i,i+1}|x\rangle,\eqno(2.3)$$
   $$C|x\rangle=\sum_{i=1}^3\delta_i(|x\rangle)(\sqrt{3-t}E_{ii})|x\rangle
+\sum_{i=1}^3\gamma_i(|x\rangle)\sqrt{t}E_{i,i+1}|x\rangle,\eqno(2.4)$$
and the matrix
$$F_{x}=\left(\begin{array}{cccccc}
 \alpha_{1}(|x\rangle) & \alpha_{2}(|x\rangle) &\alpha_{3}(|x\rangle)&\beta_{1}(|x\rangle)&\beta_{2}(|x\rangle)&\beta_{3}(|x\rangle) \\
\delta_1(|x\rangle)&\delta_2(|x\rangle)&\delta_3(|x\rangle)&\gamma_1(|x\rangle)
& \gamma_2(|x\rangle)&  \gamma_3(|x\rangle)
\end{array}\right)$$ is contractive.

Note that $\|F_x\|\leq1$ if and only if $\|F_xF_x^\dag\|\leq1 $.

In the sequel, for any unit $|x\rangle\in {\mathbb C}^3$, we  write
$|x\rangle=(|x_1|e^{i\theta_1},|x_2|e^{i\theta_2},|x_3|e^{i\theta_3})^T$.
Then $|x_1|^2+|x_2|^2+|x_3|^2=1$.

{\bf Subcase 1.} $|x_1|=|x_2|=|x_3|=\frac{1}{\sqrt{3}}$.

In Eqs.(2.3)-(2.4), by taking
$$(\alpha_1,\alpha_2,\alpha_3)=(\frac{\sqrt{3-t}}{3},\frac{\sqrt{3-t}}{3},\frac{\sqrt{3-t}}{3})\ \ \ {\rm and}\ \
(\delta_1,\delta_2,\delta_3)=(\frac{
\sqrt{3-t}c}{3},-\frac{\sqrt{3-t}c}{3},0),$$ we get
$$(\beta_1,\beta_2, \beta_3)=(\frac{ {\sqrt{t}}x_1}{3x_2},
\frac{{\sqrt{t}}x_2}{3x_3},\frac{{\sqrt{t}}x_3}{3x_1})\ \ \ {\rm
and}\ \ (\gamma_1,\gamma_2,\gamma_3)=(\frac{ {\sqrt{t}c}x_1}{3x_2},
-\frac{{\sqrt{t}}cx_2}{3x_3},0).$$ So
$\sum_{i=1}^3(|\alpha_i|^2+|\beta_i|^2)=1$,
$\sum_{i=1}^3(|\delta_i|^2+|\gamma_i|^2)=\frac{6c^2}{9}$ and
$\sum_{i=1}^3(\alpha_i\overline{\delta_i}+\beta_i\overline{\gamma_i})
=0$. It follows that
$$F_{x}F_{x}^\dag=\left(\begin{array}{cc}
 \sum_{i=1}^3(|\alpha_i|^2+|\beta_i|^2)&\sum_{i=1}^3(\alpha_i\bar{\delta_i}+\beta_i\bar{\gamma_i})\\
\sum_{i=1}^3(\bar{\alpha_i}\delta_i+\bar{\beta_i}\gamma_i)&\sum_{i=1}^3(|\delta_i|^2+|\gamma_i|^2)
\end{array}\right)=\left(\begin{array}{cc}
1& 0\\
0&\frac{6c^2}{9}\end{array}\right),$$ which implies that
$\|F_xF_x^\dag\|\leq 1\Leftrightarrow c^2\leq \frac{9}{6}.$ Hence
$$ c^2\leq 1-t\Rightarrow \|F_xF_x^\dag\|\leq 1.$$

{\bf Subcase 2.} $x_i\not=0$ for all $i=1,2,3$ and
$(|x_1|,|x_2|,|x_3|)\not=(\frac{1}{\sqrt{3}},\frac{1}{\sqrt{3}},\frac{1}{\sqrt{3}})$.

Let $r_i=|\frac{x_i}{x_{i+1}}|^2$ for  $i=1,2,3$. Then $r_i>0$,
$i=1,2,3$, and $r_1r_2r_3=1$. Take
$$(\alpha_1,\alpha_2,\alpha_3)=(\frac{\sqrt{3-t}r_1}{t+(3-t)r_1},\frac{\sqrt{3-t}r_2}{t+(3-t)r_2},\frac{\sqrt{3-t}r_3}{t+(3-t)r_3})$$
and $$(\delta_1,\delta_2,\delta_3)
=(\frac{\sqrt{3-t}r_1c}{t+(3-t)r_1},-\frac{\sqrt{3-t}r_2c}{t+(3-t)r_2},0).$$
By Eqs.(2.3)-(2.4), we get $$(\beta_1,\beta_2, \beta_3)
=(\frac{\sqrt{r_1}}{t+(3-t)r_1}e^{i(\theta_1-\theta_2)},
\frac{\sqrt{r_2}}{t+(3-t)r_2}e^{i(\theta_2-\theta_3)},
\frac{\sqrt{r_3}}{t+(3-t)r_3}e^{i(\theta_3-\theta_1)})$$and
$$(\gamma_1,\gamma_2,\gamma_3)
=( \frac{\sqrt{tr_1}c}{t+(3-t)r_1}e^{i(\theta_1-\theta_2)},
-\frac{\sqrt{tr_2}c}{t+(3-t)r_2}e^{i(\theta_2-\theta_3)},0).$$ So
$$f(\alpha_1,\alpha_2, \alpha_3)
=\sum_{i=1}^3|\alpha_i|^2+\sum_{i=1}^3|\beta_i|^2=\sum_{i=1}^3\frac{r_i}{t+(3-t)r_i},$$
$$f_{C_0}(\delta_1,\delta_2,\delta_3)=\sum_{i=1}^3|\delta_i|^2+\sum_{i=1}^3|\gamma_i|^2=\frac{r_1c^2}{t+(3-t)r_1}+\frac{r_2c^2}{t+(3-t)r_2}$$
and
$$\sum_{i=1}^3(\alpha_i\bar{\delta_i}+\beta_i\bar{\gamma_i})=\frac{r_1c}{t+(3-t)r_1}-\frac{r_2c}{t+(3-t)r_2}.$$
It follows that
$$\begin{array}{rl}F_{x}F_{x}^\dag=&\left(\begin{array}{cc}
 \sum_{i=1}^3(|\alpha_i|^2+|\beta_i|^2)&\sum_{i=1}^3(\alpha_i\bar{\delta_i}+\beta_i\bar{\gamma_i})\\
\sum_{i=1}^3(\bar{\alpha_i}\delta_i+\bar{\beta_i}\gamma_i)&\sum_{i=1}^3(|\delta_i|^2+|\gamma_i|^2)
\end{array}\right)\\
=&\left(\begin{array}{cc}
\sum_{i=1}^3\frac{r_i}{t+(3-t)r_i}& \frac{r_1c}{t+(3-t)r_1}-\frac{r_2c}{t+(3-t)r_2}\\
\frac{r_1c}{t+(3-t)r_1}-\frac{r_2c}{t+(3-t)r_2}&\frac{r_1c^2}{t+(3-t)r_1}+\frac{r_2c^2}{t+(3-t)r_2}\end{array}\right).\end{array}$$
Note that $\|F_{x}F_{x}^\dag\|\leq 1$ if and only if its maximal
eigenvalue $\lambda_{{\rm max}}\leq 1$. By a calculation, it is
easily checked that
$$\lambda_{{\rm max}}\leq 1$$ holds if and only if $$ c^2\leq \frac{1-\sum_{i=1}^3\frac{r_i}{t+(3-t)r_i}}{(1-\sum_{i=1}^3\frac{r_i}{t+(3-t)r_i})
(\frac{r_1}{t+(3-t)r_1}+\frac{r_2}{t+(3-t)r_2})+(\frac{r_1}{t+(3-t)r_1}-\frac{r_2}{t+(3-t)r_2})^2},\eqno(2.5)$$
where $r_1,r_2,r_3>0$ with $r_1r_2r_3=1$ and
$(r_1,r_2,r_3)\not=(1,1,1)$. Let
$$g(r_1,r_2,r_3)=\frac{1-\sum_{i=1}^3\frac{r_i}{t+(3-t)r_i}}{(1-\sum_{i=1}^3\frac{r_i}{t+(3-t)r_i})
(\frac{r_1}{t+(3-t)r_1}+\frac{r_2}{t+(3-t)r_2})+(\frac{r_1}{t+(3-t)r_1}-\frac{r_2}{t+(3-t)r_2})^2}.$$
Replacing  $r_i$ by $\frac{1}{r_i}$ in the above function
$g(r_1,r_2,r_3)$,  we  have
$$g(r_1,r_2,r_3)=\frac{1-\sum_{i=1}^3\frac{1}{(3-t)+tr_i}}{\sum_{i=1}^2\frac{1}{(3-t)+tr_i}
-\frac{4}{(3-t+tr_1)(3-t+tr_2)}-\frac{1}{(3-t+tr_1)(3-t+tr_3)}-\frac{1}{(3-t+tr_2)(3-t+tr_3)}}.$$
Now applying Lemma 2.4, we see that $$g(r_1,r_2,r_3)\geq 1-t$$ holds
for all positive numbers $r_1,r_2,r_3$ with $r_1r_2r_3=1$ and
$(r_1,r_2,r_3)\not=(1,1,1)$. This and Eq.(2.5) imply
$$c^2\leq
(1-t)\Rightarrow\lambda_{{\rm max}}\leq
1\Rightarrow\|F_{x}F_{x}^\dag\|\leq 1.$$

{\bf Subcase 3.} $x_1=0$ and $x_i\not=0$ for  $i=2,3$.

In this case, by Eqs.(2.3)-(2.4), one  may choose
$\beta_1=\delta_3=\gamma_1=0$, $\alpha_3=\frac{1}{\sqrt{3-t}}$,
$\beta_2=\frac{(1-\sqrt{3-t}\alpha_2)x_2}{\sqrt{t}x_3}$ and
$\gamma_2=\frac{(-c-\sqrt{3-t}\delta_2)x_2}{\sqrt{t}x_3}$. Write
$r_2=|\frac{x_2}{x_3}|^2=\frac{|x_2|^2}{1-|x_2|^2}$. Then by taking
$$\begin{array}{rl}&(\alpha_1,\alpha_2,\alpha_3,\beta_1,\beta_2,\beta_3)=
(0,\frac{\sqrt{3-t}r_2}{t+(3-t)r_2},\frac{1}{\sqrt{3-t}},0,\frac{\sqrt{t}\sqrt{r_2}}{t+(3-t)r_2}e^{i(\theta_2-\theta_3)},0)\end{array}$$
and
$$(\delta_1,\delta_2,\delta_3,\gamma_1,\gamma_2,\gamma_3)=
(0,-\frac{\sqrt{t}\sqrt{r_2}c}{t+(3-t)r_2}e^{i(\theta_2-\theta_3)},0),$$
which meet Eqs.(2.3)-(2.4), we get
$$f(\alpha_1,\alpha_2, \alpha_3)
=\sum_{i=1}^3|\alpha_i|^2+\sum_{i=1}^3|\beta_i|^2=\frac{r_2}{t+(3-t)r_2}+\frac{1}{3-t},$$
$$f_{C_0}(\delta_1,\delta_2,\delta_3)=\sum_{i=1}^3|\delta_i|^2+\sum_{i=1}^3|\gamma_i|^2=\frac{r_2c^2}{t+(3-t)r_2}$$
and
$$\sum_{i=1}^3(\alpha_i\bar{\delta_i}+\beta_i\bar{\gamma_i})=-\frac{r_2c}{t+(3-t)r_2}.$$
Hence
$$F_{x}F_{x}^\dag=\left(\begin{array}{cc}
\frac{r_2}{t+(3-t)r_2}+\frac{1}{3-t}& -\frac{r_2c}{t+(3-t)r_2}\\
-\frac{r_2c}{t+(3-t)r_2}&\frac{r_2c^2}{t+(3-t)r_2}\end{array}\right).$$
Still, by a calculation, one can easily obtain
$$\|F_{x}F_{x}^\dag\|\leq 1\Leftrightarrow c^2\leq \frac{1-\frac{r_2}{t+(3-t)r_2}-\frac{1}{3-t}}{\frac{r_2}{t+(3-t)r_2}
-\frac{r_2}{(t+(3-t)r_2)(3-t)}},\eqno(2.6)$$ where $ r_2 >0$ is any
positive number. Let
$$g(r_2)=\frac{1-\frac{r_2}{t+(3-t)r_2}-\frac{1}{3-t}}{\frac{r_2}{t+(3-t)r_2}
-\frac{r_2}{(t+(3-t)r_2)(3-t)}}.$$ A direct calculation yields that,
$$g(r_2)\geq 1-t\Leftrightarrow r_2\geq \frac{t(2-t)}{t-1}.\eqno(2.7) $$
Note that $r_2>0$ and  $\frac{t(2-t)}{t-1}<0$ as $0<t<1$. So we
always have   $r_2\geq \frac{t(2-t)}{t-1}$. Thus by Eq.(2.7), we
have proved that $g( r_2 )\geq 1-t$ holds for all positive numbers $
r_2>0$. It follows from Eq.(2.6) that
$$ c^2\leq 1-t\Rightarrow\|F_{x}F_{x}^\dag\|\leq 1.$$

{\bf Subcase 4.} $x_2=0$ and $x_i\not=0$ for  $i=1,3$.

Let $r_3=|\frac{x_3}{x_1}|^2=\frac{|x_3|^2}{1-|x_3|^2}$. By
Eqs.(2.3)-(2.4), we can choose $\beta_2=\gamma_2=0$,
$\alpha_1=\frac{1}{\sqrt{3-t}}$,
$\alpha_3=\frac{\sqrt{3-t}r_3}{t+(3-t)r_3}$,
$\beta_3=\frac{(1-\sqrt{3-t}\alpha_3)x_3}{\sqrt{t}x_1}$ and
$\gamma_3=\frac{-\sqrt{3-t}\delta_3x_3}{\sqrt{t}x_1}$. Now  take
$$\begin{array}{rl}&(\alpha_1,\alpha_2,\alpha_3,\beta_1,\beta_2,\beta_3)=
(\frac{1}{\sqrt{3-t}},0,\frac{\sqrt{3-t}r_3}{t+(3-t)r_3},0,0,\frac{\sqrt{t}\sqrt{r_3}}{t+(3-t)r_3}e^{i(\theta_3-\theta_1)})\end{array}$$
and
$$(\delta_1,\delta_2,\delta_3,\gamma_1,\gamma_2,\gamma_3)=
(\frac{c}{\sqrt{3-t}},0,0,0,0,0).$$ It follows that
$$f(\alpha_1,\alpha_2, \alpha_3)
=\sum_{i=1}^3|\alpha_i|^2+\sum_{i=1}^3|\beta_i|^2=\frac{r_3}{t+(3-t)r_3}+\frac{1}{3-t},$$
$$f_{C_0}(\delta_1,\delta_2,\delta_3)=\sum_{i=1}^3|\delta_i|^2+\sum_{i=1}^3|\gamma_i|^2=\frac{c^2}{3-t}$$
and
$$\sum_{i=1}^3(\alpha_i\bar{\delta_i}+\beta_i\bar{\gamma_i})=\frac{c}{3-t}.$$
Hence
$$F_{x}F_{x}^\dag=\left(\begin{array}{cc}
\frac{r_3}{t+(3-t)r_3}+\frac{1}{3-t}& \frac{c}{3-t}\\
\frac{c}{3-t}&\frac{c^2}{3-t}\end{array}\right).$$ Still,   one can
easily checked that
$$\|F_{x}F_{x}^\dag\|\leq 1\Leftrightarrow c^2\leq \frac{1-\frac{r_3}{t+(3-t)r_3}-\frac{1}{3-t}}{\frac{1}{3-t}
-\frac{r_3}{(t+(3-t)r_3)(3-t)}} $$ and
$$\frac{1-\frac{r_3}{t+(3-t)r_3}-\frac{1}{3-t}}{\frac{1}{3-t}
-\frac{r_3}{(t+(3-t)r_3)(3-t)}}\geq 1-t\Leftrightarrow t\geq
(t-1)r_3,$$ where $ r_3>0$ is any positive number   and   $0<t<1$.
Note that $t\geq (t-1)r_3$ as $0<t<1$ and $r_3>0$.   Thus we  see
that we still have
$$c^2\leq 1-t\Rightarrow\|F_{x}F_{x}^\dag\|\leq 1.$$

{\bf Subcase 5.} $x_3=0$ and $x_i\not=0$ for  $i=1,2$.

Let $r_1=|\frac{x_1}{x_2}|^2=\frac{|x_1|^2}{1-|x_1|^2}$. By
Eqs.(2.3)-(2.4), one may choose $\beta_3=\gamma_3=0$,
$\alpha_2=\frac{1}{\sqrt{3-t}}$,
$\beta_1=\frac{(1-\sqrt{3-t}\alpha_1)x_1}{\sqrt{t}x_2}$,
$\delta_2=\frac{-c}{\sqrt{3-t}}$ and
$\gamma_1=\frac{(c-\sqrt{3-t}\delta_1)x_1}{\sqrt{t}x_2}$. Then for
choice
$$\begin{array}{rl}&(\alpha_1,\alpha_2,\alpha_3,\beta_1,\beta_2,\beta_3)=
(\frac{\sqrt{3-t}r_1}{t+(3-t)r_1},\frac{1}{\sqrt{3-t}},0,\frac{\sqrt{t}\sqrt{r_1}}{t+(3-t)r_1}e^{i(\theta_1-\theta_2)},0,0)\end{array}$$
and
$$(\delta_1,\delta_2,\delta_3,\gamma_1,\gamma_2,\gamma_3)=
(\frac{\sqrt{3-t}r_1c}{t+(3-t)r_1},\frac{-c}{\sqrt{3-t}},0,
\frac{\sqrt{t}\sqrt{r_1}c}{t+(3-t)r_1}e^{i(\theta_1-\theta_2)},0,0),$$
we get
$$f(\alpha_1,\alpha_2, \alpha_3)
=\sum_{i=1}^3|\alpha_i|^2+\sum_{i=1}^3|\beta_i|^2=\frac{r_1}{t+(3-t)r_1}+\frac{1}{3-t},$$
$$f_{C_0}(\delta_1,\delta_2,\delta_3)=\sum_{i=1}^3|\delta_i|^2+\sum_{i=1}^3|\gamma_i|^2=\frac{r_1c^2}{t+(3-t)r_1}+\frac{c^2}{3-t}$$
and
$$\sum_{i=1}^3(\alpha_i\bar{\delta_i}+\beta_i\bar{\gamma_i})=\frac{r_1c}{t+(3-t)r_1}-\frac{c}{ 3-t }.$$
So
$$F_{x}F_{x}^\dag=\left(\begin{array}{cc}
\frac{r_1}{t+(3-t)r_1}+\frac{1}{3-t}& \frac{r_1c}{t+(3-t)r_1}-\frac{c}{ 3-t }\\
\frac{r_1c}{t+(3-t)r_1}-\frac{c}{ 3-t
}&\frac{r_1c^2}{t+(3-t)r_1}+\frac{c^2}{3-t}\end{array}\right).$$ It
is easily checked that
$$\|F_{x}F_{x}^\dag\|\leq 1\Leftrightarrow c^2\leq \frac{1-\frac{r_1}{t+(3-t)r_1}-\frac{1}{3-t}}
{(1-\frac{r_1}{t+(3-t)r_1}-\frac{1}{3-t})(\frac{r_1}{t+(3-t)r_1}+\frac{1}{3-t})
+(\frac{r_1}{t+(3-t)r_1}-\frac{1}{3-t})^2},\eqno(2.8)$$ where $r_1
>0$ is any positive number  and $0<t<1$. Let
$$g(r_1 )=\frac{1-\frac{r_1}{t+(3-t)r_1}-\frac{1}{3-t}}
{(1-\frac{r_1}{t+(3-t)r_1}-\frac{1}{3-t})(\frac{r_1}{t+(3-t)r_1}+\frac{1}{3-t})
+(\frac{r_1}{t+(3-t)r_1}-\frac{1}{3-t})^2}.$$ By a direct
calculation, one gets
$$g(r_1)\geq 1-t\Leftrightarrow 1-t^2+tr_1\geq  0. $$
Hence we always have $g(r_1)\geq 1-t$. This and Eq.(2.8) yield again
$$c^2\leq 1-t\Rightarrow\|F_{x}F_{x}^\dag\|\leq 1.$$

{\bf Subcase 6.} $x_1=x_2=0$ and $x_3\not=0$.

By Eqs.(2.3)-(2.4), we have $\beta_2=\delta_3=\gamma_2=0$ and
$\alpha_3=\frac{1}{\sqrt{3-t}}$. Then take
$$(\alpha_1,\alpha_2,\alpha_3,\beta_1,\beta_2,\beta_3)=(0,0,\frac{1}{\sqrt{3-t}},0,0,0)\ {\rm
and}\
(\delta_1,\delta_2,\delta_3,\gamma_1,\gamma_2,\gamma_3)=(0,0,0,0,0,0).
$$
We obtain $F_{x}F_{x}^\dag=\left(\begin{array}{cc}
\frac{1}{ 3-t} & 0\\
0&0\end{array}\right),$ which is contractive.

{\bf Subcase 7.} $x_1=x_3=0$ and $x_2\not=0$.

By Eqs.(2.3)-(2.4), we have $\beta_1=\gamma_1=0$,
$\alpha_2=\frac{1}{\sqrt{3-t}}$ and $\gamma_2=\frac{c}{\sqrt{3-t}}$.
Then by taking
$$(\alpha_1,\alpha_2,\alpha_3,\beta_1,\beta_2,\beta_3)=(0,\frac{1}{\sqrt{3-t}},0,0,0,0)\ {\rm
and}\
(\delta_1,\delta_2,\delta_3,\gamma_1,\gamma_2,\gamma_3)=(0,0,0,0,\frac{c}{\sqrt{3-t}},0),$$
we obtain $F_{x}F_{x}^\dag=\left(\begin{array}{cc}
\frac{1}{ 3-t} & \frac{c}{ 3-t} \\
\frac{c}{ 3-t} &\frac{c^2}{ 3-t}
\end{array}\right)=\frac{1}{ 3-t} \left(\begin{array}{cc}
1& c\\
c &c^2 \end{array}\right)$.   It is easy to check that
$$\|F_{x}F_{x}^\dag\|\leq 1\Leftrightarrow c^2\leq
(2-t). $$ So $c^2\leq 1-t$ implies $\|F_{x}F_{x}^\dag\|\leq 1$.

{\bf Subcase 8.} $x_2=x_3=0$ and $x_1\not=0$.

The case is the same as Case 7.

Thus, by combining Subcases 1-8 and applying Lemma 2.3, we have
proved that, for any matrix $C_0={\rm diag}(c,-c,0)$ with $0<c^2\leq
1-t$, the map $A\mapsto\Phi_{t,\pi}(A)-C_0AC_0^\dag$ is positive.
Then, by Theorem 1.1, we see that $W_{\Phi_{t,\pi}}$ is not optimal
whenever $l(\pi)=3$ and $0<t<1$.

The proof is finished. \hfill$\Box$

\section{Conclusions}

Every entangled state can be detected by an optimal entanglement
witness. So, it is important to construct as many as possible
optimal EWs. A natural way of constructing optimal EWs is through
NCP positive maps by Choi-Jamio{\l}kowski isomorphism
$\Phi\leftrightarrow W_\Phi$. In \cite{HLPQS}, for $0\leq t\leq n$,
 a class of new $D$-type positive maps $\Phi_{t,\pi}:
M_n({\mathbb C})\rightarrow M_n({\mathbb C})$ induced by an
arbitrary permutation $\pi$ of $(1,2,\ldots,n)$ was constructed,
where $\Phi_{t,\pi}$ is defined by
$$\Phi_{t,\pi}(A)=(n-t)\sum_{i=1}^nE_{ii}AE_{ii}+t\sum_{i=1}^nE_{i,\pi(i)}AE_{i,\pi(i)}^\dag-A. \eqno(3.1)$$
It was shown in \cite{HLPQS} that $\Phi_{t,\pi}$ in NCP positive if
and only if $0<t\leq\frac{n}{l(\pi)}$. In \cite{QH}, by using
Theorem 1.1, we proved that $W_{\Phi_{1,\pi}}$ is optimal if
$l(\pi)=n$ and $\pi^2\not= {\rm id}$. But it is not clear that
whether or not there exist other optimal $W_{\Phi_{t,\pi}}$s. We
guess there are no.

{\bf  Conjecture.} {\it For $n\geq 3$, $W_{\Phi_{t,\pi}}$ is an
optimal entanglement witness if and only if $t=1$, $l(\pi)=n$ and
$\pi^2\not= {\rm id}$.}

The case $n=2$ is simple. It is easily checked that
$W_{\Phi_{t,\pi}}$ is   optimal if and only if $t=1$ and $l(\pi)=2$.
Note that, $\pi^2=$id if $n=2$.

The present note gives  an affirmative answer to the above
conjecture for the case $n=3$.


\end{document}